\documentclass[aps,pra,twocolumn,groupedaddress,nofootinbib,showpacs]{revtex4-1}

\usepackage{graphicx}
\usepackage[labelfont=small]{subcaption}
\usepackage{amsmath}

\graphicspath{{figures/}}
\begin{document}

\title{Superstatistical velocity distributions of cold trapped ions in molecular dynamics simulations }

\author{I. Rouse}
\affiliation{Department of Chemistry, University of Basel, Basel, Switzerland}
\author{S. Willitsch}

\affiliation{Department of Chemistry, University of Basel, Basel, Switzerland}

\date{\today}

\begin{abstract}
We present a realistic molecular-dynamics treatment of laser-cooled ions in radiofrequency ion traps which avoids previously made simplifications such as modeling laser cooling as a friction force and combining individual heating mechanisms into a single effective heating force. Based on this implementation, we show that infrequent energetic collisions of single ions with background gas molecules lead to pronounced heating of the entire ion ensemble and a time-varying secular ensemble temperature which manifests itself in a superstatistical time-averaged velocity distribution of the ions. The effect of this finding on the experimental determination of ion temperatures and rate constants for cold chemical reactions is discussed.
\end{abstract}


\maketitle

%%%%% INTRODUCTION

\section{Introduction}
The study of the properties of atoms and molecules at low temperatures has been greatly advanced by the development of particle traps, with applications in many fields of physics and chemistry \cite{willitsch08b,raab87a,ockeloen13a,gerlich92a}. Radiofrequency (RF) ion traps in particular have proven to be a useful tool for experiments in quantum information and the study of fundamental chemical processes due to their ability to confine a wide range of atomic and molecular ions prepared at a low temperature by laser or sympathetic cooling \cite{haeffner08a, willitsch12a}. The fluorescence generated during laser cooling can be captured with a CCD camera allowing the ions to be imaged. The resulting image can be compared to the results of molecular dynamics (MD) simulations of the trajectories of the trapped ions allowing the number and the temperature of the trapped ions to be inferred \cite{williams90a,prestage91a,zhang07a, ostendorf06a,tong10a, okada10a}. In reaction experiments, this procedure is often adopted to probe the variation of the composition of the ion ensemble as a function of reaction time in order to determine reaction rate constants\cite{roth06b, tong10a, bell09a, okada13a}. 

Two significant approximations are typically made in MD simulations in order to minimize the computation time which scales quadratically with the number of trapped ions due to the pairwise Coulomb force. The first of these is to implement the effects of laser cooling as a friction force neglecting the process of stimulated emission and the saturation of the force with respect to intensity and ion velocity \cite{stenholm86a}. The second is the replacement of the various heating processes acting on the ions with a single effective heating force and fitting the strength of this force to experimental data \cite{roth05a,zhang07a, okada10a,hall13a,mokhberi14a}. This approach neglects the underlying physical details of the heating sources and, in particular, replaces rare events which lead to strong heating with a continuous, weaker heating process. This converts the system from one which samples a broad range of temperatures into one in which the ions remain at an essentially fixed temperature after reaching equilibrium with only small displacements from the mean. 

Alternative descriptions of laser cooling have previously been used ranging from stochastic photon scattering \cite{wuebbena12a}, to modeling the radiative dynamics of each particle explicitly as a two-level system in a laser beam \cite{marciante10a}, to tracking the quantum state of each particle throughout the simulation including multiple hyperfine levels \cite{jonsell06a,svensson08a}. All of these methods have the advantage of directly introducing the heating of the particles caused by photon recoil and fluctuations in the scattering rate, whereas these must be included in an effective heating term if a frictional force is used.

Building on this work, we show that these two approximations can be problematic simplifications which lead to pronounced misrepresentations of the dynamics and thermal properties of the trapped ions. We introduce an MD implementation of a realistic force model for trapping, laser cooling and the salient heating mechanisms which is directly based on experimental parameters without relying on the above two approximations. We show the effects of the breakdown of the friction model of laser cooling at ion velocities of relevance to experiments. Moreover, we demonstrate that collisions of only a single ion of the ensemble with a background gas molecule can lead to significant deviations of the secular velocities of all trapped ions from a Maxwell-Boltzmann distribution. We compare the present theoretical results to experimental data finding excellent agreement and invoke superstatistics to provide a simple explanation for the appearance of the non-Maxwellian behavior. Finally, we discuss the relevance of our results to the determination of reaction rate constants and put them into context with other recent theoretical work on the collisional dynamics of trapped ions \cite{devoe09a, chen14a}.

%%%%% BACKGROUND

\section{Background}

%%%%% RF TRAPS

\subsection{Radiofrequency ion traps}

The trapping of ions through electric fields requires a time-varying component, since the Laplace equation of electrostatics forbids the formation of a three-dimensional trap using only static fields \cite{major05a}. In a quadrupole RF ion trap, confinement is achieved through combining RF and static electric fields in such a way that charged particles can be trapped in all three dimensions simultaneously \cite{major05a}. The dynamics of a single ion in a quadrupole RF trap are governed by the Mathieu equations of motion,

\begin{equation}
\frac{d^2 u_i}{dt^2} = \frac{\Omega_\text{RF}^2 }{4}( 2q_i\cos(\Omega_\text{RF} t) -a_i)  u_i,
\end{equation}

where $u_i,i\in\{x,y,z\},$ are the ion coordinates, $q_i$ and $a_i$ are dimensionless parameters depending on the trap geometry and operating conditions, and $\Omega_\text{RF}$ is the RF frequency \cite{major05a}.

When $q_i$ and $a_i$ are both much smaller than 1, the motion of the ion can be decomposed into two components: a slow ``secular'' low frequency motion and a fast oscillating ``micromotion'' at the driving RF frequency \cite{major05a, berkeland98a}. The velocities $v$ of the ions corresponding to the secular motion can be calculated by averaging their total velocities $v_{tot}$ including micromotion over one RF period \cite{zhang07a}. For multiple trapped ions, the instantaneous distribution of secular velocities $f_v(v)$ is found to be Maxwellian in good approximation \cite{schiffer00a}:
\begin{equation}\label{maxwellDist}
f_v (v) = 4 \pi v^2  \left(  \frac{m}{2 \pi k_\text{B} T} \right)^{\frac{3}{2}} e^{-\frac{m v^2}{2 k_{\text{B}} T}},
\end{equation}
so that a secular temperature $T$ can be assigned to the ion ensemble. Here, $m$ refers to the ion mass and $k_\text{B}$ is Boltzmann's constant.

%%%%% LASER COOLING

\subsection{Laser cooling}
\label{sec:lc}
Cooling of the ions is implemented through the scattering of photons from a laser beam slightly red detuned from an atomic resonance (Doppler laser-cooling) \cite{foot05a}. Unlike in the optical-molasses cooling technique for neutral atoms, there is often only one cooling laser present in ion-cooling experiments. The motional degrees of freedom  of all ions are strongly coupled through the Coulomb interaction and so cooling one degree of freedom is sufficient to cool the entire ensemble. 

The rate $R_{scatt} (v_{tot})$ of photons scattered by an ion through interaction with a near-resonant laser beam of detuning $\delta_L$ and intensity $I$ on a transition with natural linewidth $\Gamma$ and saturation intensity $I_{sat}$ is given by  \cite{foot05a}:

\begin{equation} \label{eq:rscatt}
R_{scatt}(v_{tot}) =  \frac{\Gamma}{2} \frac{I/I_{sat}}{1 + I/I_{sat} + 4 (\delta_L + \mathbf{k} \cdot \mathbf{v_{tot}})^2 / \Gamma^2}.
\end{equation}
The corresponding scattering force 
\begin{equation}
F_{scatt} = \hbar k R_{scatt} \label{eq:fscatt}
\end{equation}
can be expanded in a Taylor series around $v_{tot}=0$ which is terminated after the linear term yielding a radiation pressure term $F_0$ and a friction force $F_f = -\beta_v v_{tot}$ \cite{foot05a}. This linear approximation of the laser-cooling force has frequently been employed in previous MD treatments of trapped ions \cite{zhang07a, tong10a, okada10a}. $F_0$ acts to slightly displace the ions in the trap, and $F_f$ acts as a velocity-dependent term removing kinetic energy from the system. The randomly-directed spontaneous emission of excited ions and random fluctuations in the number of photons absorbed per unit time both lead to a random walk of the ion velocity \cite{foot05a}. This is equivalent to a heating of the ions with a rate
\begin{equation}\label{eq:laserHeatingRate}
 \dot T =  \frac{2}{3} \frac{\hbar^2 k^2}{m k_\text{B}}   R_{scatt}.
\end{equation}

In the case of an RF ion trap, the laser cooling acts to reduce the secular temperature. At sufficiently low secular temperatures, the ion ensemble undergoes a phase transition to an ordered state called a Coulomb crystal \cite{willitsch08b,willitsch12a} . The micromotion, on the other hand, is driven by the applied RF field and so cannot be cooled completely, but can be reduced by careful control of shape and location of the Coulomb crystal \cite{hall13a}.

%%%%% FORCE MODEL

\section{Force model}
A realistic MD simulation of ions in a RF trap must take into account the range of forces experienced by the ions. These include the trapping force and Coulomb repulsion between the ions as well as the laser cooling responsible for lowering their temperature. In addition to these, a number of heating processes counteract the effect of cooling such that an ion cloud reaches a finite temperature, typically in the range 10-100 mK \cite{willitsch12a}. These processes can be broadly divided into four types: heating due to interaction with the laser, heating due to collisions of ions with residual background gas, heating due to ion-ion collisions, and heating due to experimental imperfections (e.g., machining imperfections, electronic noise and patch potentials). 

Here, the force acting on an ion $i$ was represented as follows:

\begin{equation}
F_i = - \nabla U(x,y,z,t) + F_{Coulomb} + F_{background} + F_{scatt},
\end{equation}

where $U(x,y,z,t)$ is the time-dependent trapping potential and $F_{Coulomb}$ is the sum of pairwise Coulomb forces acting between the ions. These two terms also implicitly account for heating by ion-ion collisions \cite{chen13b}. $F_{background}$ is a term representing elastic collisions with the background gas, and $F_{scatt}$ is the force arising from interaction of the ions with near-resonant light. Heating due to experimental imperfections was neglected in the present treatment (see below).

\subsection{Trapping potential}
In the present MD implementation, the trapping potential was formulated as \cite{house08a}:
\begin{equation}
U(x,y,z,t) = \frac{\Omega_{RF}^2 m}{8}   \sum_{i=x,y,z} (a_{i} -2q_{i}\cos(\Omega_{RF} t) ) u_i^2 .
\end{equation}
All simulations performed in this work use the potential of the surface-electrode chip trap described in \cite{mokhberi15a} defined by $\Omega_{RF} = 8 \times 2 \pi$ MHz, $q_x=0.0824,q_y=-0.0806,q_z=-3\times10^{-4}$ and $a_x=-16\times10^{-4},a_y=11\times10^{-4},a_z=4\times10^{-4}$, where the $a_i$ and $q_i$ parameters were derived from numerical trapping potentials \cite{mokhberi15a}.

%%%%% ION-LASER INTERACTION

\subsection{Ion-laser interaction} \label{section:laserModel}

We assume that a trapped ion can be described as a two-level atom in the weak-binding approximation such that the standard treatment of a free atom in a laser field can be applied. Moreover, we assume that the population of the excited state is given by the steady-state solution to the optical Bloch equations (OBE) \cite{foot05a} as a function of laser intensity and ion velocity, i.e., that changes in the ions internal state occur much faster than in its position or velocity. To convert this to a form suitable for use in MD simulations, we write the transfer of population between the upper and lower state in the form of Einstein rate equations (neglecting coherences between states) with the rate of change of population $P_i$ of the upper state $(i=2)$ given by

\begin{equation} \label{eq:rateEquation}
\frac{d P_2}{dt} = \gamma P_1 - (\gamma+\Gamma)P_2 ,
\end{equation}

where $\Gamma$ is the natural linewidth of the transition corresponding to the rate coefficient of spontaneous emission and $\gamma$ is the rate coefficient of absorption and stimulated emission (assuming equal degeneracies in both states). By equating the steady-state solution of Eq.~(\ref{eq:rateEquation}) to the steady-state solution of the OBE  \cite{foot05a} we obtain:

\begin{equation}
P_{2} =  \frac{\gamma}{2\gamma +\Gamma} = \frac{\Omega^2/4}{(\delta_L + \mathbf{k} \cdot \mathbf{v_{tot}})^2 + \Omega^2/2 + \Gamma^2/4}. \label{eq:obe}
\end{equation}

From Eq. (\ref{eq:obe}), an expression for $\gamma$ can be found:
 
\begin{equation} \label{eq:gammaEquation}
\gamma = \frac{\Gamma}{2} \frac{I}{I_{sat}} \frac{\Gamma^2}{ \Gamma^2 + 4(\delta_L + \mathbf{k} \cdot \mathbf{v_{tot}})^2}, 
\end{equation}

where the substitution $\Omega^2 = \frac{\Gamma^2}{2}  \frac{I}{ I_{sat}}   $ has been made. The probability that an atom in the excited state undergoes spontaneous emission during a short period of time $\delta t$ is equal to $\Gamma \delta t$. Additionally, the probability of stimulated emission from the excited state or absorption in the ground state are given by $\gamma \delta t$. Thus, the continuous rate equations for an ensemble of atoms can be converted to a probabilistic model for individual atoms. This can be implemented in an MD simulation with a timestep of $\delta t$, under the condition that $\delta t << 1/\gamma$,$1/\Gamma$ such that not more than one absorption or emission event occurs per timestep. 

\begin{figure}[bt] 
\centering
\includegraphics[width=0.5\textwidth]{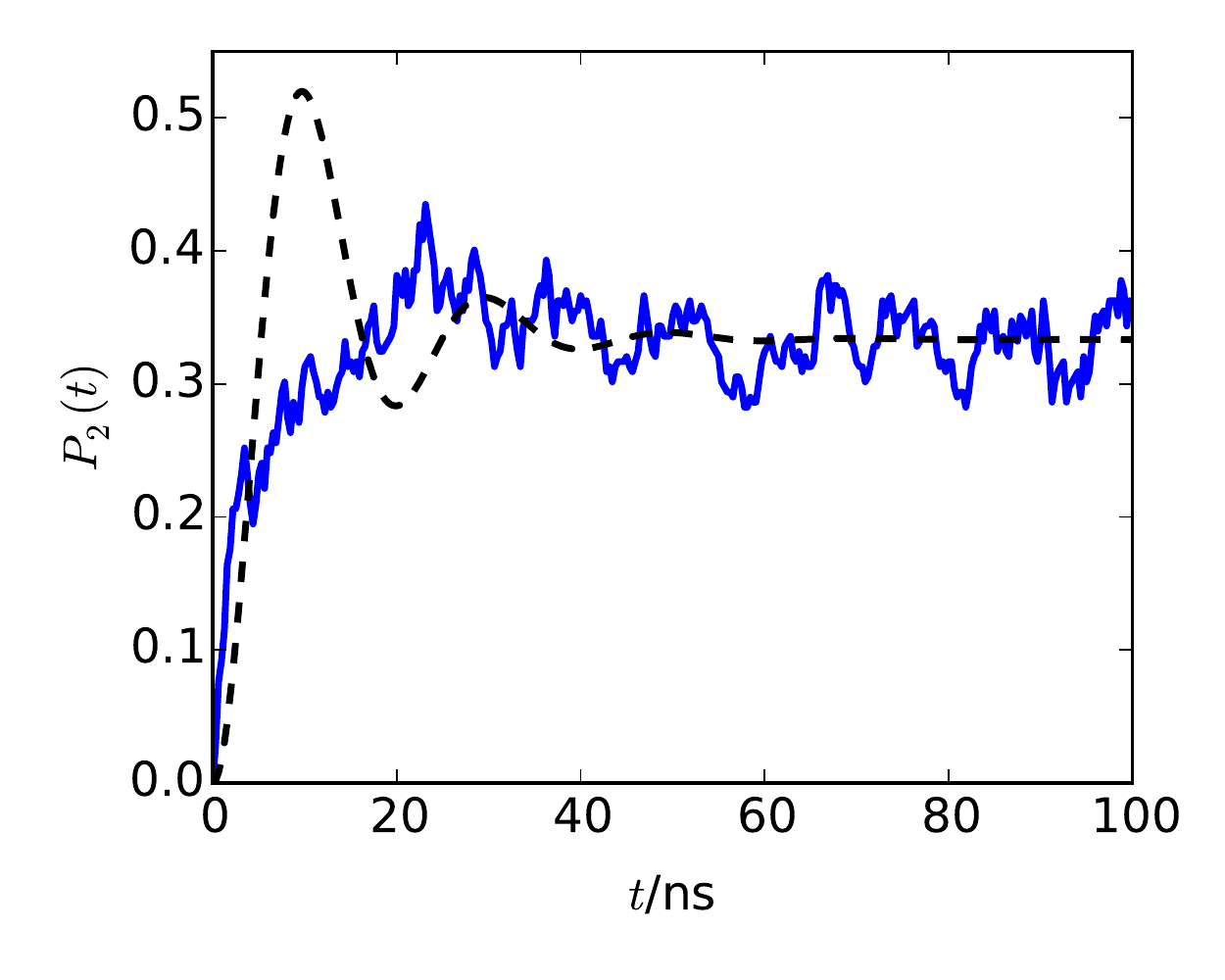}
\caption{Population $P_2$ of the upper state of a two-level atom as a function of time $t$ given by the solution to the optical Bloch equation (black dashed line) and fraction of ions in the upper state of an ensemble of 262 Ca$^+$ ions undergoing discrete transitions between a lower and upper state according to the``state-tracking" model described in the text (blue solid line). }
\label{fig:rabiAndDiscrete}
\end{figure}

In the initialization of the simulations, each ion is assigned a boolean state variable (0=ground state, 1=excited state). During each timestep, the probability of the ion changing to the other state is calculated, and this value is compared to a random number generated in the interval [0,1) to determine if a transition between the two states takes place during the timestep. When a transition occurs, the state variable of the ion is updated, and a momentum kick of magnitude $\hbar k$ is applied to the ion in the appropriate direction for absorption and stimulated emission and in a random direction for spontaneous emission. Figure~\ref{fig:rabiAndDiscrete} shows the time-dependent fraction of Ca$^+$ ions in the excited state in a simplified simulation in which the motion of the particles is neglected ($v=0$ for all ions) performed at $I=10 I_{sat}$ and $\delta_L=\Gamma$ (blue line). The black dashed line shows the corresponding solution of the OBE. We assume laser cooling on the $4s ^2 S_{1/2} \to 4p ^2 P_{1/2}$ transition in Ca$^+$ with spectroscopic parameters taken from \cite{nist14a}. It can be seen that the population obtained with the present probabilistic treatment quickly converges to the same value as obtained from the OBE. The present approach does not capture the population oscillations obtained from the OBE shortly after the beginning of the excitation. However, on the long time scales of ms considered in the present study, these deviations at the beginning of the dynamics are insignificant.

\begin{figure}[tb] 
\centering
\includegraphics[width=0.5\textwidth]{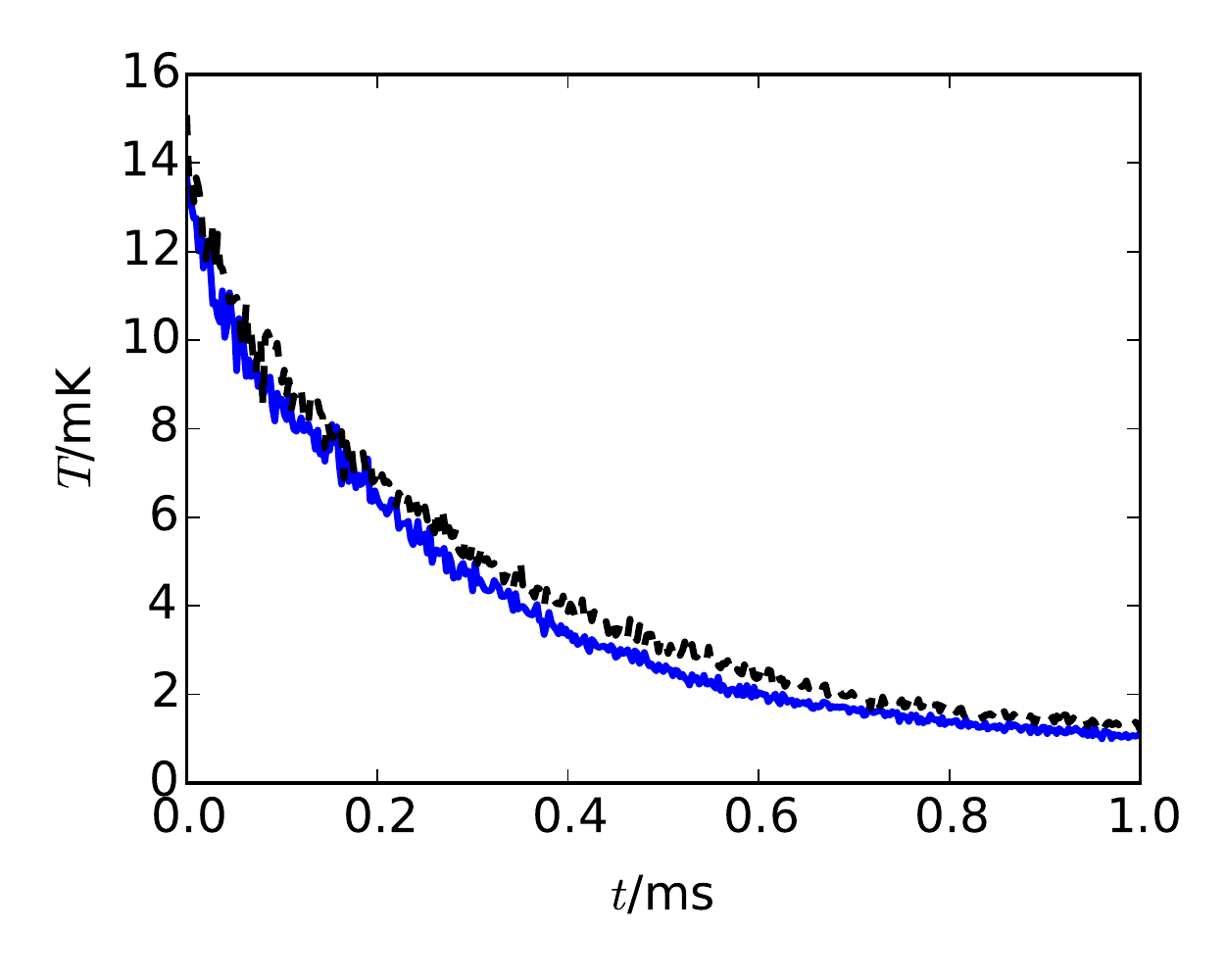}
\caption{Secular temperature $T$ as a function of time $t$ of a system of 262 ions undergoing laser cooling at an intensity of 0.2$I_{sat}$ and a detuning  $\delta_L = \Gamma$ for the friction-force (black dashed line) and state-tracking (blue solid line) models of laser cooling. }
\label{fig:coolingCurveLowIntensity}
\end{figure}

In a full MD simulation, the transition probabilities become velocity dependent as a consequence of the Doppler shift in Eq.~(\ref{eq:obe}). An ion moving towards the source of a red-detuned laser beam absorbs more photons than one moving away resulting in a net force due to the applied momentum kicks. This leads to an overall cooling effect, as can be seen in Fig.~\ref{fig:coolingCurveLowIntensity}, and automatically generates the stochastic heating force resulting from laser-cooling discussed in Sec. \ref{sec:lc}. To allow for a comparison with the friction-force model of Sec. \ref{sec:lc}, the laser intensity in this simulation was set far below saturation, $I=0.2 I_{sat}$, such that the friction-force model using the heating rate of Eq.~(\ref{eq:laserHeatingRate}) is applicable. In this low-intensity regime, it can be seen that the two models give very similar results. In Sec. \ref{section:Validation}, we will discuss the differences between the two models at higher laser intensities more closely related to those used in experiments.

It should be noted that the present two-level model neglects the existence of metastable states such as the $^2 D_{3/2}$ state in Ca$^+$ which experimentally requires a repumper laser to prevent the accumulation of population in this state. Consequently, it also neglects the effects of a possible coherent population trapping which, however, can be suppressed by a suitable choice of laser detunings \cite{hall13b}. 

%%%%% BACKGROUND GAS

\subsection{Collisions with background gas}
Even at the ultrahigh vacuum of lower than 10$^{-9}$ mbar usually used in cold-ion experiments, there is a significant rate of elastic collisions of ions with residual background gas, typically H$_2$. These collisions are primarily caused by the long range interaction between the ion and the induced dipole of the neutral molecule, and so the collision rate constant can be approximated by Langevin theory \cite{gioumousis58a}:
\begin{equation}
k_{el} = 2 \pi n_n \sqrt{\frac{\alpha'_n e^2}{4 \pi \epsilon_0 \mu}} ,
\end{equation}
with background particle density $n_n$, polarizability volume $\alpha'_n$ and the reduced mass of the collision $\mu$. As already shown in Ref.~\cite{zhang07a}, a collision of an ion with a background gas molecule imparts momentum to the ions ejecting it from the crystal for a number of oscillation periods until it is recooled. Kinetic energy is transferred from the hot ion to the remaining ions in the crystal through ion-ion collisions resulting in an increase of the secular temperature followed by laser-recooling to equilibrium. In our simulations, the assumption is made that the collisions are purely elastic following a hard-sphere model. This is a reasonable approximation for an ion interacting with a weakly polarizable neutral particle. Although alternatives could certainly be considered, we emphasize that the salient effects described later in this paper do not depend on the choice of collision model.  For the crystal with 262 ions of Fig. \ref{fig:coolingCurveLowIntensity} interacting with H$_2$ molecules at a temperature of 300K and partial pressure of 10$^{-9}$ mbar, an average of one collision event per 30ms takes place. A single simulation covering this period of time took approximately 10 hours. Thus, it would take a prohibitive computation time to average over the large number of collisions which occur on the timescales of typical experiments (seconds to minutes). One method of overcoming this problem is to increase the rate and decrease the size of the momentum kicks imparted in the collisions so that the average heating rate remains the same \cite{zhang07a}. In this way, a steady-state is reached in the simulations in only a few milliseconds of simulation time. This approach, however, does not accurately reproduce the underlying dynamics of the crystal as it creates a system which remains close to equilibrium temperature with only minor deviations which is accurate only for large crystals in which collisions are frequent \cite{zhang07a}. The effect of rarer background gas collisions is to cause a sharp rise in the temperature followed by a slow recooling to equilibrium. In Section \ref{section:backgroundGas}, we will describe an approach which preserves the essential features of the realistic heating mechanism and cooling dynamics while remaining computationally tractable.

%%%%% IMPERFECTIONS

\subsection{Experimental imperfections}
Ions in an RF trap are sensitive to heating from a number of potential imperfections of the trap, such as phase differences between the RF electrodes, anomalous ion heating, the formation of patch potentials on the electrodes, and Johnson noise \cite{naini11a}. These depend on the specific experimental environment and can vary on a daily basis. Attempting to include them all explicitly in a simulation would be impractical. It has been shown that in big traps these effects are typically small even for relatively large clouds of a thousand ions \cite{zhang07a}. At present our simulations do not account for these effects. Judging from the comparison of our simulation results with experimental data in Sec. \ref{sec:val} below, we conclude that they are negligible for the trap and conditions considered here.

%%%%% NUMERICS

\subsection{Numerical implementation}
Due to its numerical stability with respect to treating oscillatory motions and its computational simplicity, the synchronized form of the leapfrog algorithm was used to integrate the equations of motion for the trapped ions \cite{skeel97a}. The code was implemented in the MD frameworks ProtoMOL and OpenMM \cite{protomol04a,openMM13a}. Simulations performed with 1000 ions on OpenMM using single GPU acceleration (nVidia GeForce \textregistered GTX 650) proved 4x faster than the ProtoMOL code running on 4 CPU cores (Intel \textregistered Xeon \textregistered CPU E5-2687W).

%%%%% RESULTS

\section{Results and discussion}

%%%%% VALIDATION OF MD MODEL

\subsection{Validation of the MD model} \label{section:Validation}
\label{sec:val}
\begin{figure}[bt] 
\centering
%\begin{subfigure}[b]{0.5\textwidth}
%\caption{~}
%\includegraphics[width=\textwidth]{laserModelsRates.pdf}
%\label{fig:modelCompareRates}
%\end{subfigure}
% 
%\begin{subfigure}[b]{0.5\textwidth}
%\caption{~}
%\includegraphics[width=\textwidth]{laserModelsTEqs.pdf}
%
%\label{fig:modelCompareEquilibrium}
%\end{subfigure}
\includegraphics[width=0.45\textwidth]{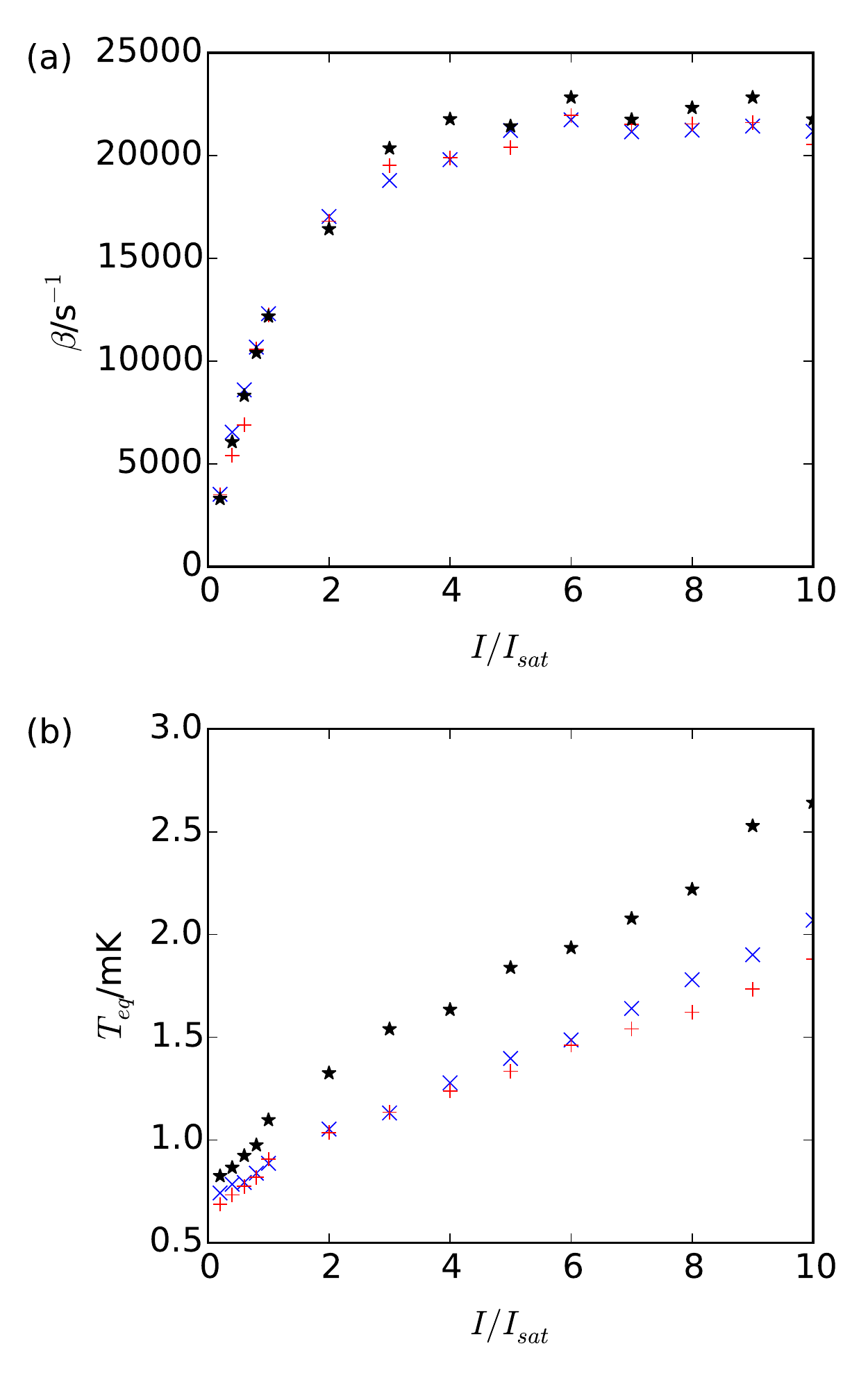}
\caption{(a) Cooling rate $\beta$ and (b) equilibrium temperature $T_{eq}$ of a cloud of 262 Ca$^+$ ions using the friction-force model (black $\star$), the present state-tracking model (blue $\times$) and the state-tracking model without stimulated emission (red $+$) at a laser intensity of 9$I_{sat}$ and a detuning $\delta_L = \Gamma$. See text for details.}
\label{fig:laserCompare}
\end{figure}

To validate our ``state-tracking'' model of laser cooling, a number of simulations investigating the cooling from a temperature of 10 mK to equilibrium were performed at a fixed detuning and varying intensity. Additional simulations were performed using the friction model of laser cooling with the heating term described by Eq.~(\ref{eq:laserHeatingRate}), and also with the state-tracking model with stimulated emission neglected (as in Ref. \cite{marciante10a}). Exponential decay curves were fitted to the data obtained in these simulations allowing extraction of the cooling rate  $\beta$  and the equilibrium temperature $T_{eq}$ as a function of the laser intensity, see Figure~\ref{fig:laserCompare}. It can be seen that $\beta$ is approximately equal for all three models with respect to $I/I_{sat}$, but $T_{eq}$ differs at high intensities because of different saturation behaviour.

\begin{figure}[tb] 
\centering
\includegraphics[width=0.45\textwidth]{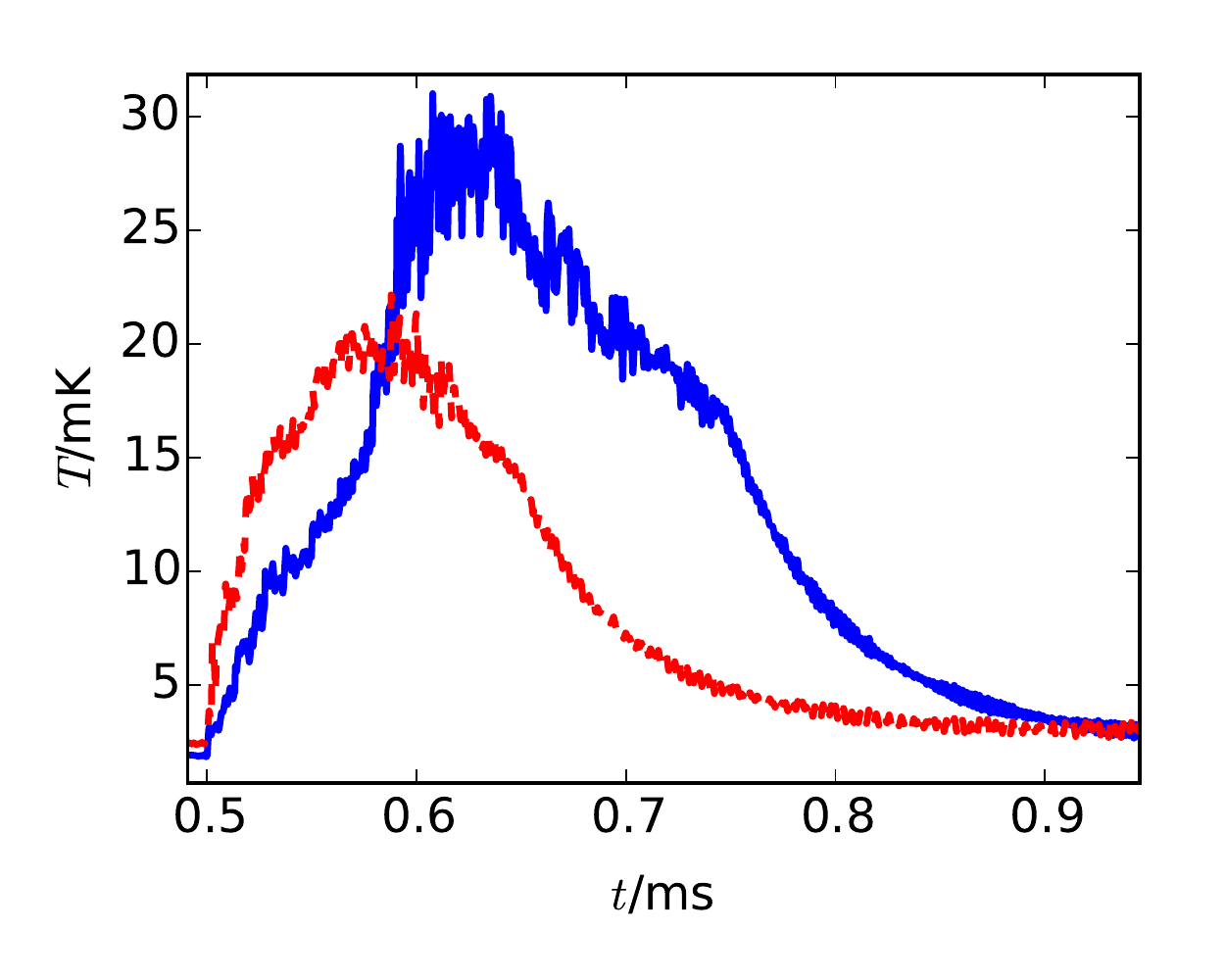}
\caption{The secular temperature $T$ as a function of time $t$ of an ion crystal of 262  Ca$^+$ ions following a collision of an ion with a H$_2$ molecule using state-tracking (blue solid line) and friction-force (red dashed line) models of laser cooling averaged over 20 simulations. The friction model yields an unrealistically fast cooling rate resulting from a breakdown of the linear approximation to the scattering force.}
\label{fig:kickRecooling}
\end{figure}

Typically, the majority of ions in the trap remain at low secular velocities for which the linear expansion inherent in the friction force applies. However, after a head-on Ca$^+$ - H$_2$ collision at a mean relative velocity of 1775~m~s$^{-1}$, the calcium ion is accelerated to a velocity of 170~m~s$^{-1}$, for which the friction model is no longer valid. The force applied to an ion at this velocity by the friction term can be up to two orders of magnitude larger than the maximum scattering force for a near-optimum detuning of $\Gamma$. This results in an unrealistically fast recooling to equilibrium, as can be seen in Fig.~\ref{fig:kickRecooling}. The present state-tracking model does not rely on the friction approximation and so allows the role of collisions to be more accurately investigated.

\begin{figure}[bt] 
\centering
%\begin{subfigure}[b]{0.5\textwidth}
%\caption{~}
%\includegraphics[width=\textwidth]{ca_262_v2_exp}
%\end{subfigure}
%
%\begin{subfigure}[b]{0.5\textwidth} 
%\caption{~}
%\includegraphics[width=\textwidth]{ca_262_frictionSim}
%\end{subfigure}
%
%\begin{subfigure}[b]{0.5\textwidth} 
%\caption{~}
%\includegraphics[width=\textwidth]{ca_262_bmean}
%\end{subfigure}
\includegraphics[width=0.45\textwidth]{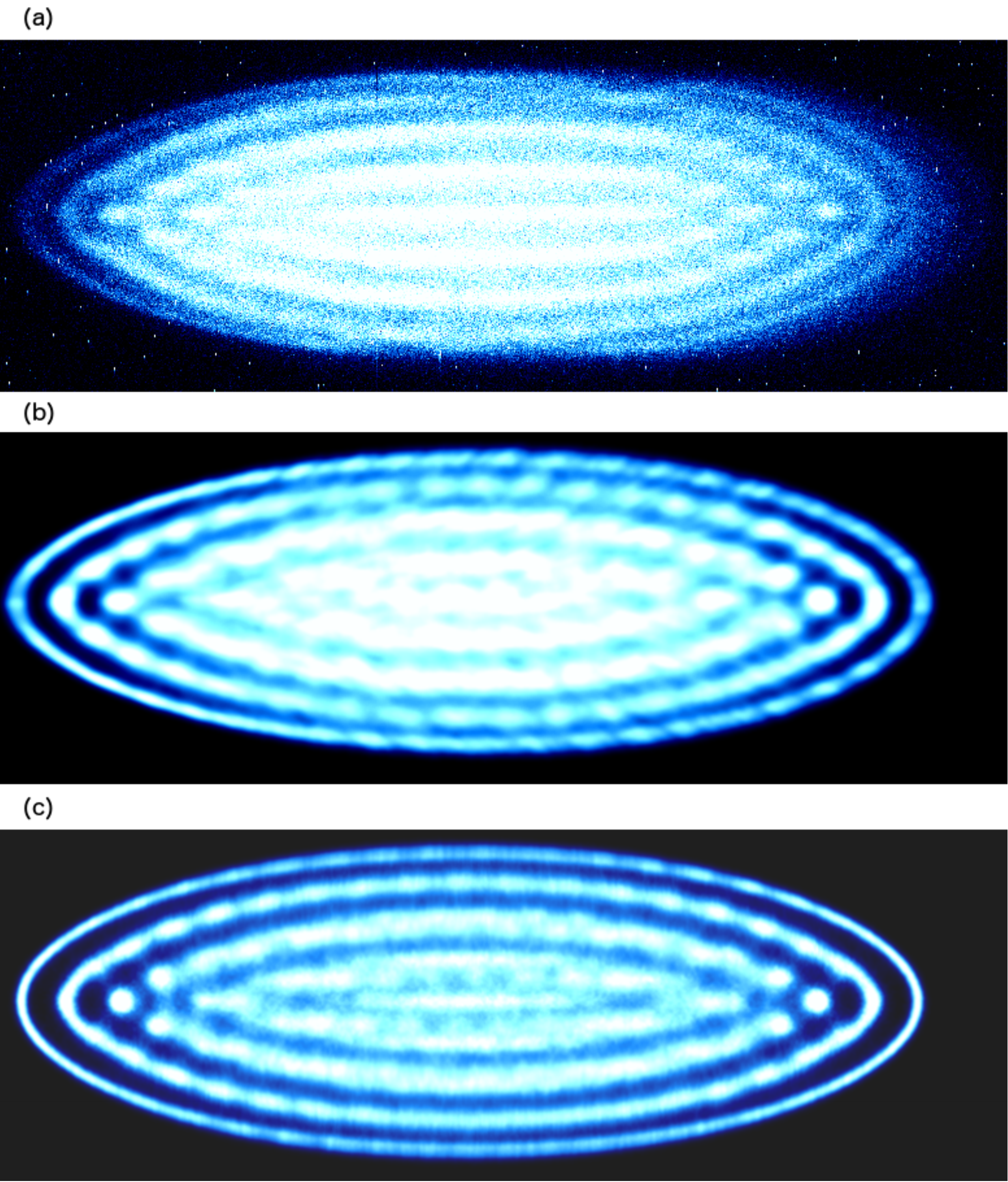}
\caption{(a) Experimental and (b) simulated CCD image of 262 Ca$^+$ ions in the six-wire trap as previously reported in \cite{mokhberi14a}. (c) Simulated image of 262 ions based on the present MD implementation described in the text.}
\label{fig:armoCrystal}
\end{figure}

In order to confirm that the present MD framework successfully reproduces experimental results, simulations of Coulomb crystals obtained in a surface-electrode RF ion trap used previously \cite{mokhberi14a, mokhberi15a} were performed. The laser intensity and detuning were set to the experimental values of $I \approx 9 I_{sat}$ and $\delta_L \approx 13 \Gamma$. The effect of collisions with background gas molecules at room temperature was simulated by applying a kick to a single ion at the mean collisional velocity and impact parameter at a fixed point in time (see Sec. \ref{section:backgroundGas}). Simulations were carried out for a time interval equal to the mean period between collisions with background gas molecules. We obtained simulated images in a good agreement with experiment and previous simulations (Fig.~\ref{fig:armoCrystal}) confirming the validity of the present MD approach. 

%%%%% COLLISIONS

\subsection{Collisions with background gas} \label{section:backgroundGas}

Assuming that the ion motions are strongly coupled, the temperature following an increase at a time $t_0$ is approximated by an exponential decay to equilibrium \cite{zhang07a,marciante10a,chen13b}: 

\begin{equation} \label{eq:timeTempSingleExp}
T(t) = T_{eq} + \Delta T(v_c,b) \exp(-\beta (t-t_0)).
\end{equation}

Here, $\Delta T(v_c,b)$ is the increase in temperature from $T_{eq}$ due to a collision occurring with a collisional velocity $v_c$ and impact parameter $b$. This expression assumes that collisions take place sufficiently infrequently such that the system recools to equilibrium in between which is the case here. Since both $\beta$ and $T_{eq}$ are independent of the strength of the collision, they can be extracted from a set of trial simulations performed at a constant collision strength. Averaging over ten simulations, the best fit parameters were found to be $\beta = 151 $~s$^{-1}$ and $T_{eq}$ = 7.78~mK.

\begin{figure}[tb] 
\centering
\includegraphics[width=0.45\textwidth]{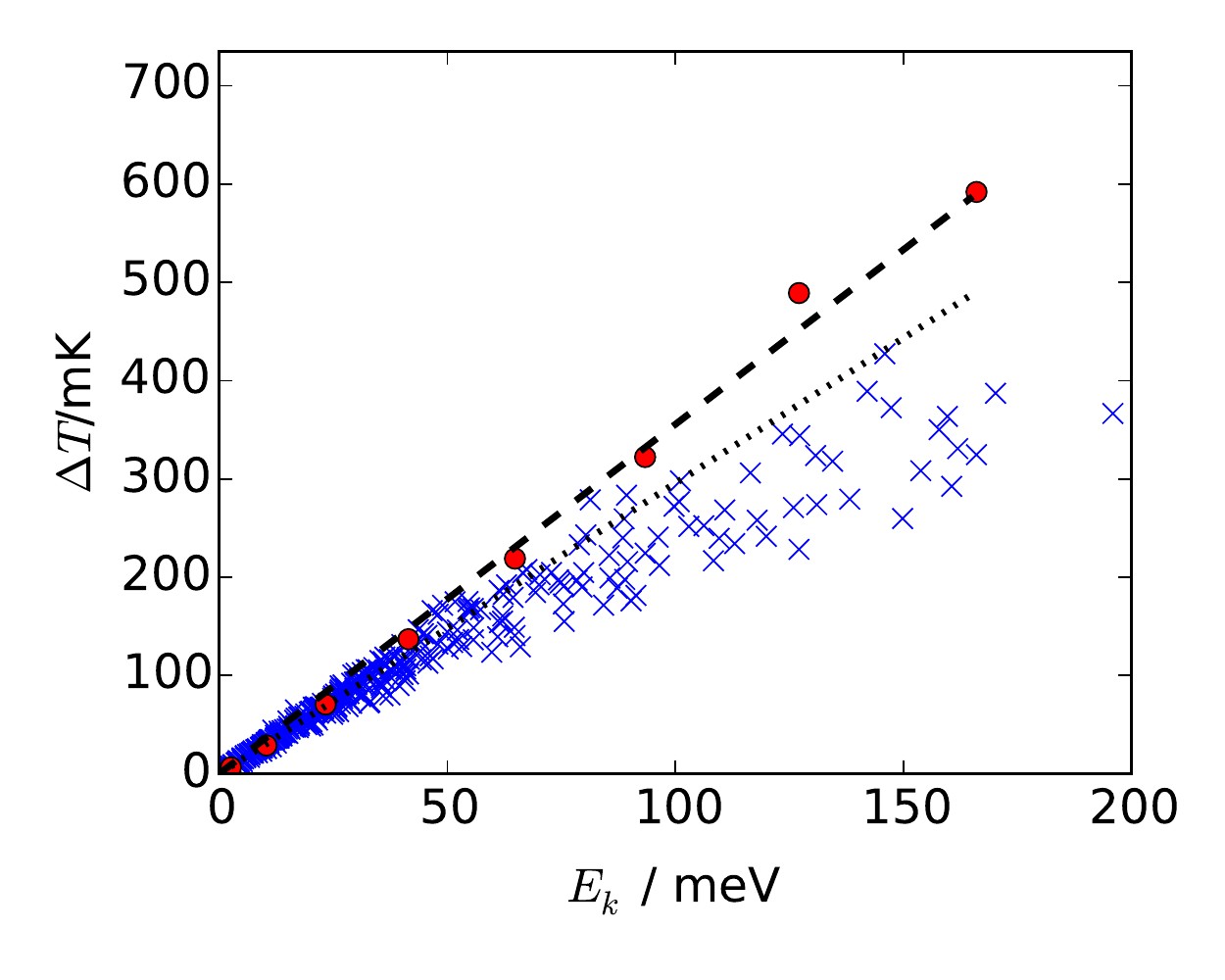}
\caption{Increase in secular temperature $\Delta T$ of a Coulomb crystal of 262 Ca$^+$ ions following a collision with a H$_2$ molecule at impact parameter $b=0$ sampled over $\approx 350$ simulations (blue crosses) as a function of the kinetic energy $E_k$ of the colliding H$_2$ molecule. The dotted line is a best fit to indicate the deviation from the linear trend at high $E_k$. Red circles indicate additional simulations performed with the laser cooling switched off after the collision, and the dashed line is a line of best fit to these data. See text for discussion.}
\label{fig:tempVelKick}
\end{figure}

To establish values for $\Delta T(v_c,b)$, a total of $\approx$ 350 simulations were performed to sample over the range of collision velocities at an H$_2$ temperature of 300~K at impact parameter $b=0$. The values of $\Delta T(v_c,b=0)$ thus obtained are plotted in Fig.~\ref{fig:tempVelKick} as a function of the kinetic energy of the colliding H$_2$ molecules. Assuming an instantaneous energy transfer between the ions and that equipartition of energy applies, the temperature increase of the crystal should be a linear function of the kinetic energy transferred in the initial collision with the background gas molecule \cite{moriwaki92a,baba02c, chen14a}. We observe this to be the case when the laser cooling is switched off following the collision, but not when the cooling remains active (see Fig.~\ref{fig:tempVelKick}).

\begin{figure}[bt] 
\centering
%\begin{subfigure}[b]{0.45\textwidth}
%\caption{~}
%\includegraphics[width=\textwidth]{3000LaserTemp.pdf}
%\end{subfigure}
%\begin{subfigure}[b]{0.45\textwidth} 
%\caption{~}
%\includegraphics[width=\textwidth]{3000LaserVel.pdf}
%\end{subfigure}
\includegraphics[width=0.45\textwidth]{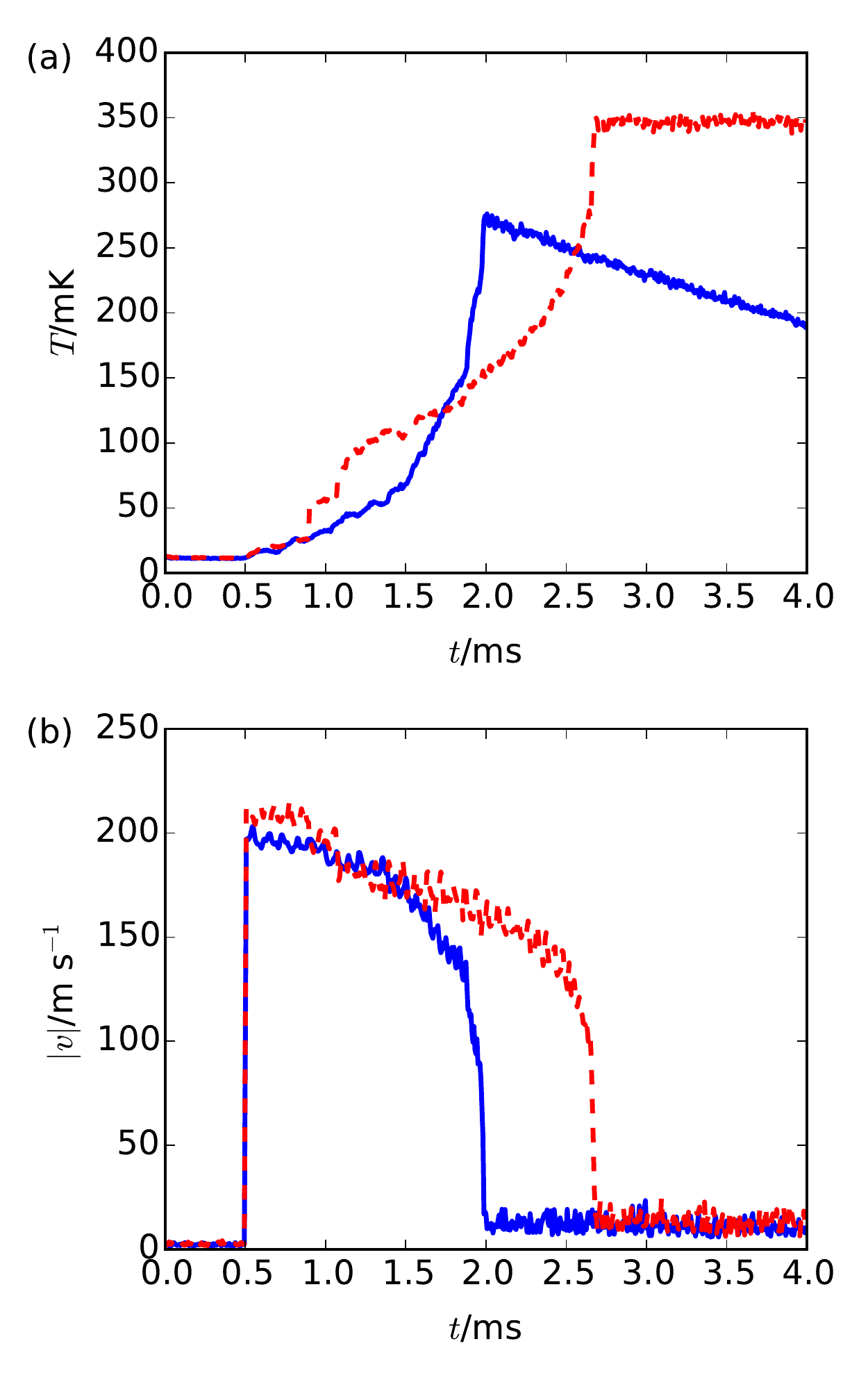}
\caption{ (a) Secular temperature $T$ of a Coulomb crystal of 262 Ca$^+$ ions following the collision of a single ion with a H$_2$ molecule at a velocity of 3000 m s$^{-1}$ and impact parameter $b=0$ with (blue solid line) and without (red dashed line) laser cooling applied. (b) Magnitude of the secular velocity $|v|$ of the ejected ion for these two cases. See text for details.}
\label{fig:highLowTempVel}
\end{figure}

This phenomenon appears to be caused by the fact that at higher velocities, the transfer of energy from the ejected ion to the crystal is not instant and instead occurs over an extended period of time as shown in Fig.~\ref{fig:highLowTempVel}. As previously observed, the ejected ion moves on a large orbit in the trap transferring energy to the colder ions only through infrequent collisions when passing through the centre \cite{zhang07a}. Laser cooling causes additional loss of kinetic energy reducing the radius of the orbit. After the ion reaches a low enough energy at which it can no longer escape from the crystal, it quickly equilibrates through collisions with the other ions. In Fig.~\ref{fig:highLowTempVel}, the re-capture of the ejected ion by the crystal occurs at approximately 2~ms in the presence of a cooling laser, and is visible as a sharp maximum in the temperature of the crystal (Fig. \ref{fig:highLowTempVel} (a)) and steep drop in the velocity of the ejected ion (Fig~\ref{fig:highLowTempVel} (b)). Without laser cooling, recapture occurs later and the overall increase in temperature of the crystal is also larger. From these results we conclude that the nonlinear trend in Fig.~\ref{fig:tempVelKick} is caused by the laser cooling process dissipating energy during the extended time between the collision and the re-thermalization of the ion which also results in a lower temperature increase of the crystal.

For a full characterization of the collision dynamics, it is necessary to consider collisions at all possible impact parameters. The average temperature increase due to collisions is then given by:
\begin{equation} \label{eq:deltaTVB}
\overline{\Delta T} =  \int \int  \Delta T(v_c,b) f_{v_c}(v_c) f_b(b) \text{d} b  \text{d} v_c ,
\end{equation}
where $f_{v_c}(v_c)$ and $f_b(b)$ are the distributions of collision velocities and impact parameters, respectively. Evaluation of this expression as detailed in Appendix \ref{ap:imp} yields $\overline{\Delta T}$ = $56.3\pm0.9$ mK for 262 laser-cooled Ca$^+$ ions colliding with room-temperature H$_2$ background gas molecules. 

Thus, a time- and collision-strength-averaged temperature $\overline{T}$ can be calculated:

\begin{equation}
\bar T = \frac{1}{\Delta t} \int_{t_0}^{t_0+\Delta t} T_{eq} + \overline{\Delta T} e^{-\beta t} dt,
\label{eq:tbar}
\end{equation}

where the averaging length $\Delta t$ is taken to be the period between collisions $\Delta t=1/k_{el}$. Using $\overline{\Delta T}$, $\beta$ and $T_{eq}$ extracted from the simulations, the mean temperature of the experimental crystal shown in Fig.~\ref{fig:armoCrystal} (a) was found to be $\overline{T}=20\pm$1~mK.

In Ref. \cite{mokhberi14a}, the same crystal was assigned a temperature of $T = 23$~mK, obtained through use of the frequent, weak collision model of Ref. \cite{zhang07a} producing the image shown in Fig.~\ref{fig:armoCrystal} (b). The difference between these results is attributed to the large temperature fluctuations in the current model which were previously neglected.

%%%%% SUPERSTATISTICS

\subsection{Superstatistical velocity distributions}
\label{sec:stat}

\begin{figure}[bt] 
\centering
\includegraphics[width=0.45\textwidth]{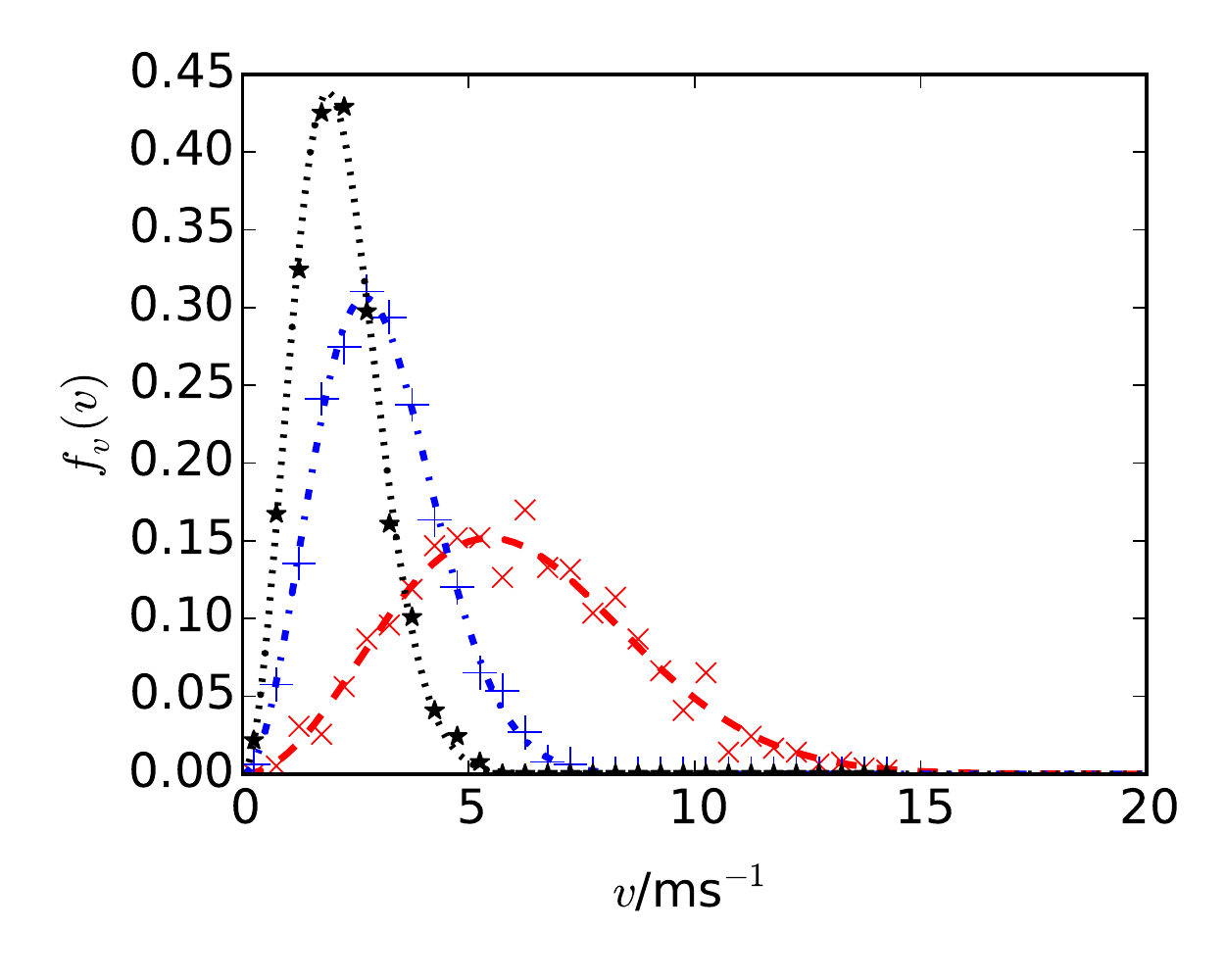}
\caption{Instantaneous secular-velocity distributions $f_v(v)$ of the crystal of 262 Ca$^+$ ions at times t=2.5~ms (red $\times$), 15~ms (blue +) and 27.5~ms (black $\star$) after a collision with a H$_2$ molecule.  Maxwell-Boltzmann distributions fitted to the numerical results at these times are shown as  dashed, dash-dotted and dotted lines, respectively. The data points are averaged over six iterations of the simulation and represent a histogram bin of width $\Delta v=0.5$~m~s$^{-1}$.}
\label{fig:velHistTimeDist}
\end{figure}

The effects of the different collision models can be investigated by sampling the secular-velocity distribution of the ions over the period $\Delta t$. For the model using frequent weak kicks to all ions, the velocity distribution is Maxwellian for a fixed temperature, as would be expected for a sample in thermal equilibrium. The realistic present model using infrequent, energetic collisions of a single ion with a single background gas molecule does not lead to a system in thermal equilibrium at a fixed temperature. The instantaneous velocity distributions of the ions at a certain point in time are found to be approximately Maxwellian, but change with time, see Fig.~\ref{fig:velHistTimeDist}. This reflects the considerably faster timescale for the energy redistribution within the crystal than for the cooling of the entire ensemble.

\begin{figure}[b] 
\centering
\includegraphics[width=0.45\textwidth]{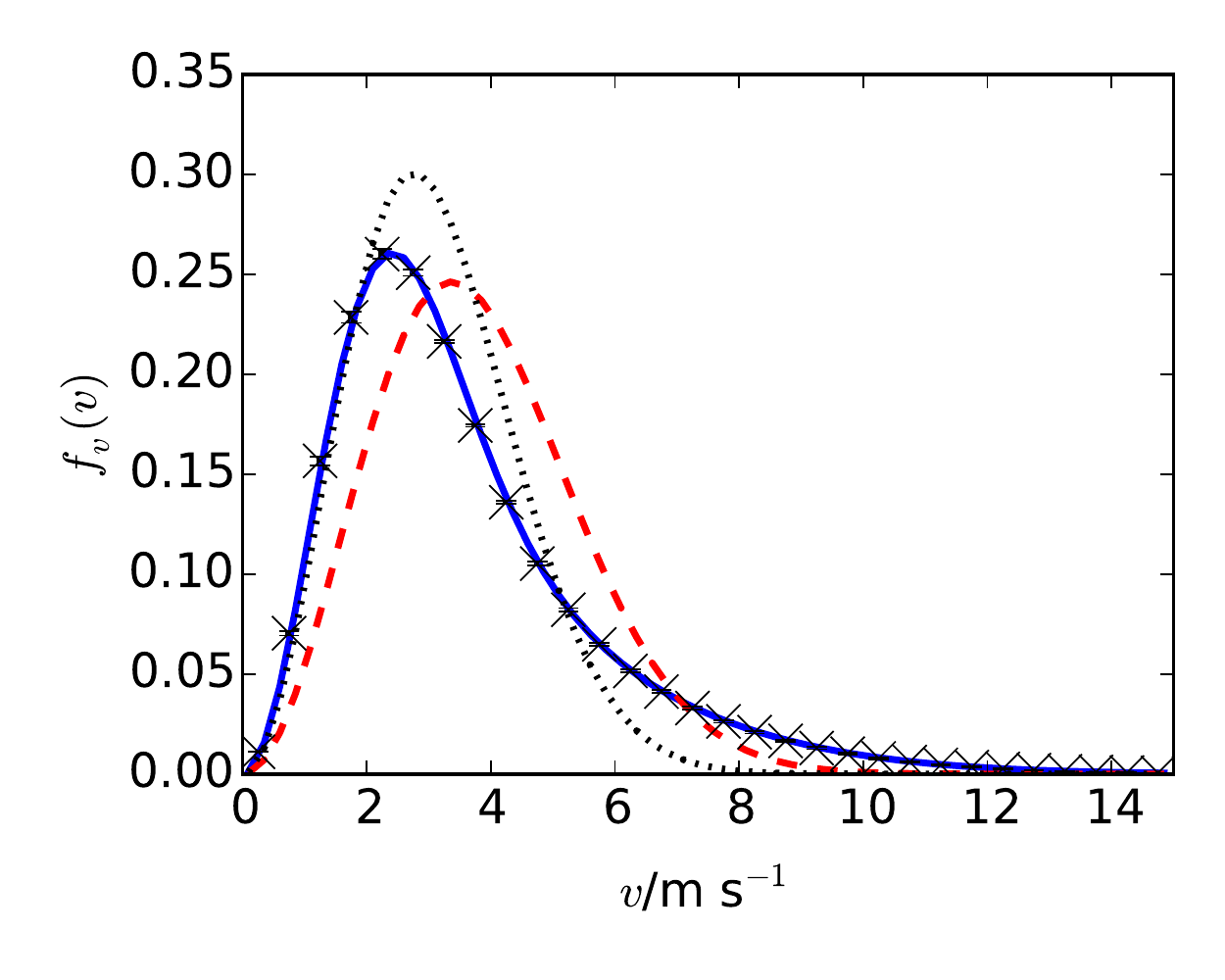}
\caption{Velocity distribution of 262 Ca$^+$ ions in the period of time between two collisions from Eq.~(\ref{eq:velDistIntegrand}) (solid blue line) and the numerical distribution obtained by sampling a simulation over this interval (crosses). The red dashed line is a Maxwell-Boltzmann distribution at the average temperature $\overline{T} = 26$~mK, and the black dotted line is a Maxwell-Boltzmann distribution  at $\hat T=18$~mK obtained from a fit to the numerical data. In the simulation, a H$_2$ velocity of 1775~m~s$^{-1}$ and impact parameter $b=0$ were assumed. }
\label{fig:ionVelDist}
\end{figure}

To simplify the analysis, we assume that collisions occur at fixed time intervals $\Delta t$, that the temperature rise is instant following a kick, and that the system recools to equilibrium in between. Thus, only the effects of a single collision need to be considered at a time, and the secular-velocity distribution found when sampling over $\Delta t$ can be written as a time average of the instantaneous thermal distributions taking into account the time-varying temperature for a given temperature rise $\Delta T$:

\begin{equation} \label{eq:velDistIntegrand}
f_v(v \vert \Delta T ) = \frac{1}{\Delta t}\int^{t_0 + \Delta t}_{t_0} 4  \pi v^2 \sqrt{\left(\frac{m}{2\pi k_{\text{B}} T(t)}\right)^3} e^{\frac{-mv^2}{2 k_{\text{B}} T(t)}} \text{d}t ,
\end{equation}
where the left-hand side has been written as $f_v(v \vert \Delta T ) $ to emphasize that this represents the distribution following a specified rise in the temperature $\Delta T$. This integral may be evaluated numerically for an arbitrary time-dependent temperature $T(t)$ and can be solved analytically for a temperature of the form given by Eq.~(\ref{eq:timeTempSingleExp}) (see Appendix~\ref{ap:timeAvDist}). We compare the result of this analytical solution to the distribution obtained numerically for simulations performed at a collision velocity $v_c=1775$~m~s$^{-1}$ and an impact parameter $b=0$ in Fig.~\ref{fig:ionVelDist}.  It can be seen that the distribution has a longer tail at high velocities than a standard Maxwell-Boltzmann distribution due to the periods of high temperature immediately following a collision, and is in very good agreement with the values calculated using Eq.~(\ref{eq:velDistIntegrand}) with $T(t)$ given by Eq. (\ref{eq:timeTempSingleExp}).

The two Maxwell-Boltzmann distributions plotted in Fig.~\ref{fig:ionVelDist}  correspond to the time-averaged temperature (Eq.~(\ref{eq:tbar})) for $v_c=1775$~m~s$^{-1}$, $b=0$ of $\overline{T}=26$~mK (red dashed line) and a temperature of $\hat T = 18$~mK (black dotted line) obtained from a fit of a Maxwell-Boltzmann distribution to the numerical data (crosses). Neither correctly describe the ion velocities -- the distribution for $T=\overline{T}$ overpredicts the peak velocity and for $T=\hat T$ the high-velocity tail is lost. For this system, the temperature fluctuations are significant enough that a single static temperature cannot accurately describe the distribution of ion velocities.

The Cartesian components of the secular velocity $v_i, \ i=x, y,z$, can also be sampled over the course of a simulation producing the distribution:

\begin{equation} \label{eq:velDistComponentIntegrand}
f_{v_i}(v_i~\vert\Delta T) = \frac{1}{\Delta t}\int^{t_0 + \Delta t}_{t_0}  \sqrt{\frac{m}{2\pi k_{\text{B}} T(t)}} e^{\frac{-mv^2}{2 k_{\text{B}} T(t)}} \text{d}t 
\end{equation}
as shown in Fig.~\ref{fig:tsallisFig} (see Appendix~\ref{ap:timeAvDist} for an analytical solution to this integral). Similar heavy-tailed distributions have previously been observed experimentally for atoms in an optical lattice, and in simulations of ions undergoing buffer-gas cooling \cite{douglas06a, devoe09a,zipkes11a,chen14a}. In these cases the velocity distributions are generally a good fit to a Tsallis function. Indeed, the data shown in Fig. \ref{fig:tsallisFig} can be fit to a Tsallis distribution of the form used in \cite{devoe09a},

\begin{equation} \label{eq:tsallisDeVoe}
f_T (x) = \frac{T_0}{ (1 + (x/\sigma)^2/n)^n } ,
\end{equation}

with excellent accuracy. The fit yields a width of the distribution $\sigma=2.34$ m s$^{-1}$ and an exponent $n=2.72$. Fig. \ref{fig:tsallisFig} also shows the solution of the integral Eq.~(\ref{eq:velDistComponentIntegrand})  (see Appendix \ref{ap:timeAvDist} for the analytic expression). It can be seen that this representation is in excellent agreement with the numerical data and the Tsallis function.

\begin{figure}[b] 
\centering
\includegraphics[width=0.45\textwidth]{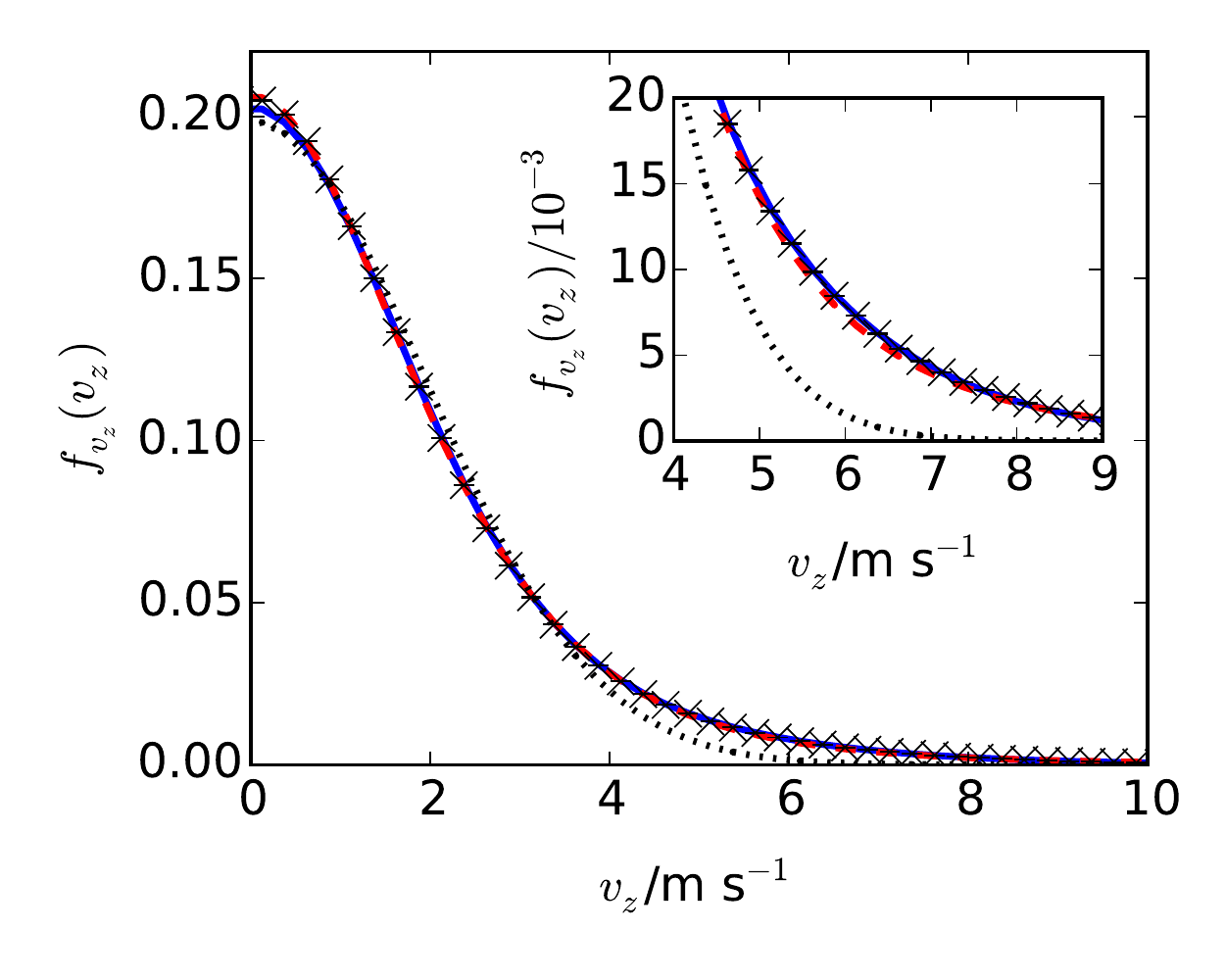}
\caption{Distribution of the axial component $v_z$ of the secular velocity (crosses) sampled during a simulation over 30ms following a collision event with a H$_2$ molecule with $v=1775$ m~s$^{-1}$ at impact parameter $b=0$. The red dashed line represents a fit to a Tsallis function and the black dotted line to a Gaussian distribution. The blue solid line is a Maxwell-Boltzmann distribution averaged over a time-varying temperature, see text for details. The inset shows the heavy-tailed behaviour of the distributions at high velocities.}
\label{fig:tsallisFig}
\end{figure}

The present results can be understood within the framework of the superstatistics of driven non-equilibrium systems \cite{beck03a}. The time average over an exponentially decaying temperature performed in Eq.~(\ref{eq:velDistIntegrand}) is mathematically equivalent to an integration over a temperature distribution $ f_T(T \vert \Delta T) \sim 1/(T-T_{eq})$, as demonstrated in Appendix~\ref{ap:timeAvDist}.  It has been shown previously that the low-energy limit of a Maxwell-Boltzmann distribution averaged over fluctuating temperature is a Tsallis distribution \cite{beck01a, beck03a}, and so the fact that it represents a good fit to the data simply reflects that the ions remain in this low energy regime. Appendix \ref{ap:timeAvDist} also provides an analytical form for the superstatistical parameter $q_s = 1 + 1/n$ \cite{beck03a}. At $q_s = 1$, the system follows Maxwell-Boltzmann statistics, and values greater than 1 indicate a greater deviation from Maxwellian behavior. A numerical investigation of this parameter revealed that over a wide range of equilibrium temperatures (1-40 mK) and values of $\Delta T$ (10-200 mK), $q_s$ is maximized for $\beta \Delta t \approx 4$, corresponding to the situation in which the crystal has just recooled to equilibrium before the next collision occurs. For much faster ($\beta \to \infty$) or much slower ($\beta \to 0$) cooling, $q_s \to 1$ and the Maxwellian limit is recovered. $\beta$, and therefore $q_s$, can be adjusted by varying the laser-cooling parameters, i.e., laser detuning and intensity.

\begin{figure}[b] 
\centering
\includegraphics[width=0.5\textwidth]{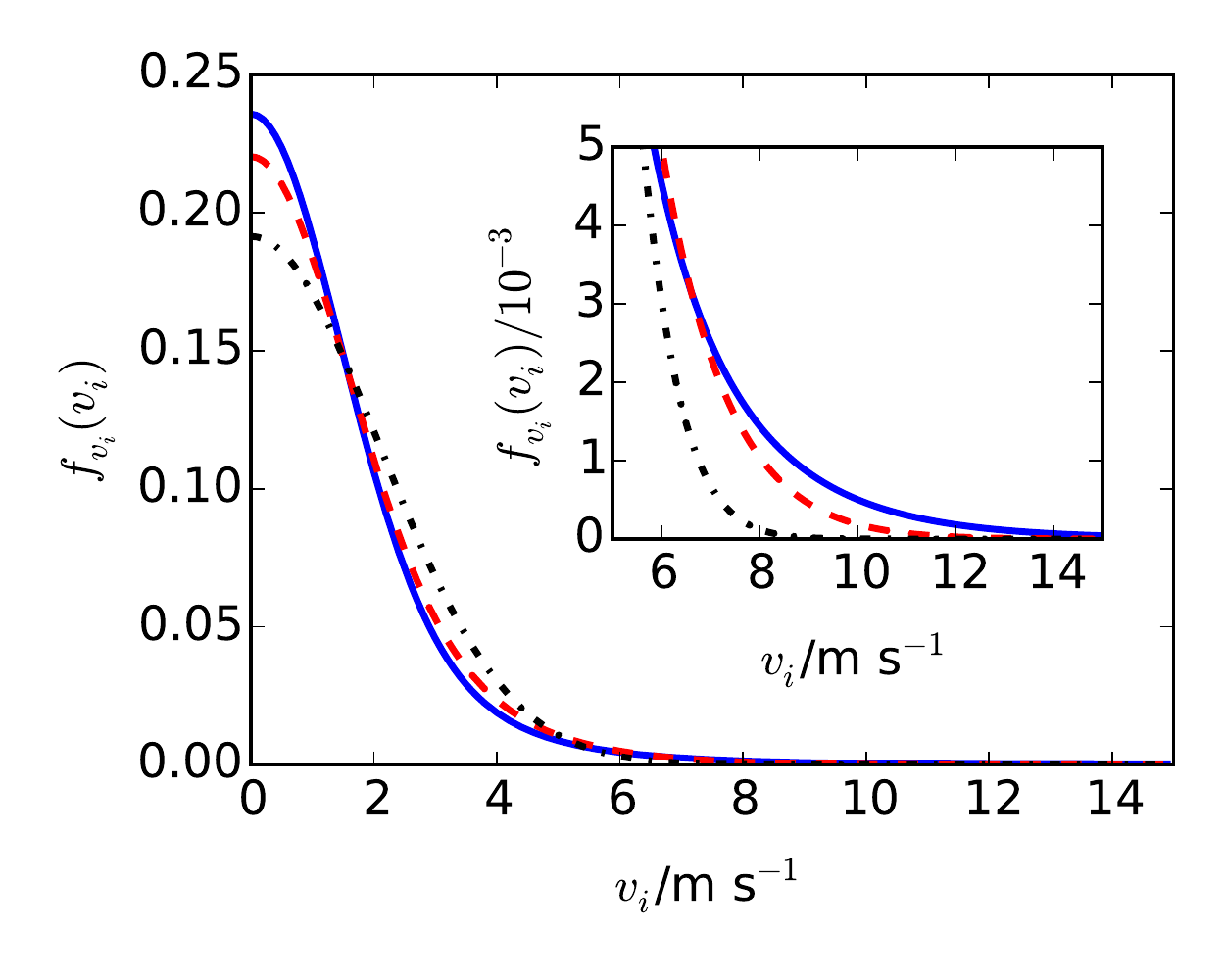}
\caption{Superstatistical velocity distributions $f_{v_i}(v_i)$ for a crystal of 262 Ca$^+$ ions. The blue solid line shows the results of averaging a Maxwell-Boltzmann velocity distribution over a varying temperature induced by collisions with a range of impact parameters $b$ and velocities $v_c$ causing increases in the temperature $\Delta T(v_c,b)$ followed by a slow return to equilibrium. The red dashed line is calculated using the same averaging procedure but treating all collisions as causing the mean increase in temperature $\overline{\Delta T}$=56.3mK. The black dash-dotted line is a Maxwell-Boltzmann distribution for a fixed temperature equal to the mean $\overline{T}$=20mK. The inset shows the heavy-tailed behavior of the superstatistical distributions.}
\label{fig:timeAverageKickAverage}
\end{figure}

The discussion above applies to the time interval following a single collision of fixed strength and hence a known value of $\Delta T$. Averaging over the distribution of collision velocities and impact parameters yields
\begin{equation}
f_{v_i}(v_i) = \int^\infty_0   f_{v_i}(v_i~\vert\Delta T(v_\bot))   f_{v_\bot}(v_\bot)~\text{d} v_\bot, \label{eq:distav}
\end{equation}
where $v_\bot$ is the normal collision velocity (see Appendix \ref{ap:imp}). Using the results obtained previously for $\Delta T$, Eq. (\ref{eq:distav}) was numerically integrated. The result is plotted in Fig.~\ref{fig:timeAverageKickAverage} and compared to both the distribution $f_{v_i}(v_i~\vert \overline{\Delta T}) $ obtained using Eq.~(\ref{eq:velDistComponentIntegrand}) with $\Delta T = \overline{\Delta T} =56.3$~mK and the Maxwell-Boltzmann distribution at the mean temperature $\overline{T}$ given by Eq. (\ref{eq:tbar}). It can be seen that averaging over all collisions in this manner leads to a distribution with an even more pronounced tail compared to  $f_{v_i}(v_i~\vert \overline{\Delta T})$, since this now includes the effects of the most energetic collisions. 

%%%%% REACTIONS

\subsection{Consequences for studies of cold chemistry}

\begin{figure}[tb] 
\centering
\includegraphics[width=0.5\textwidth]{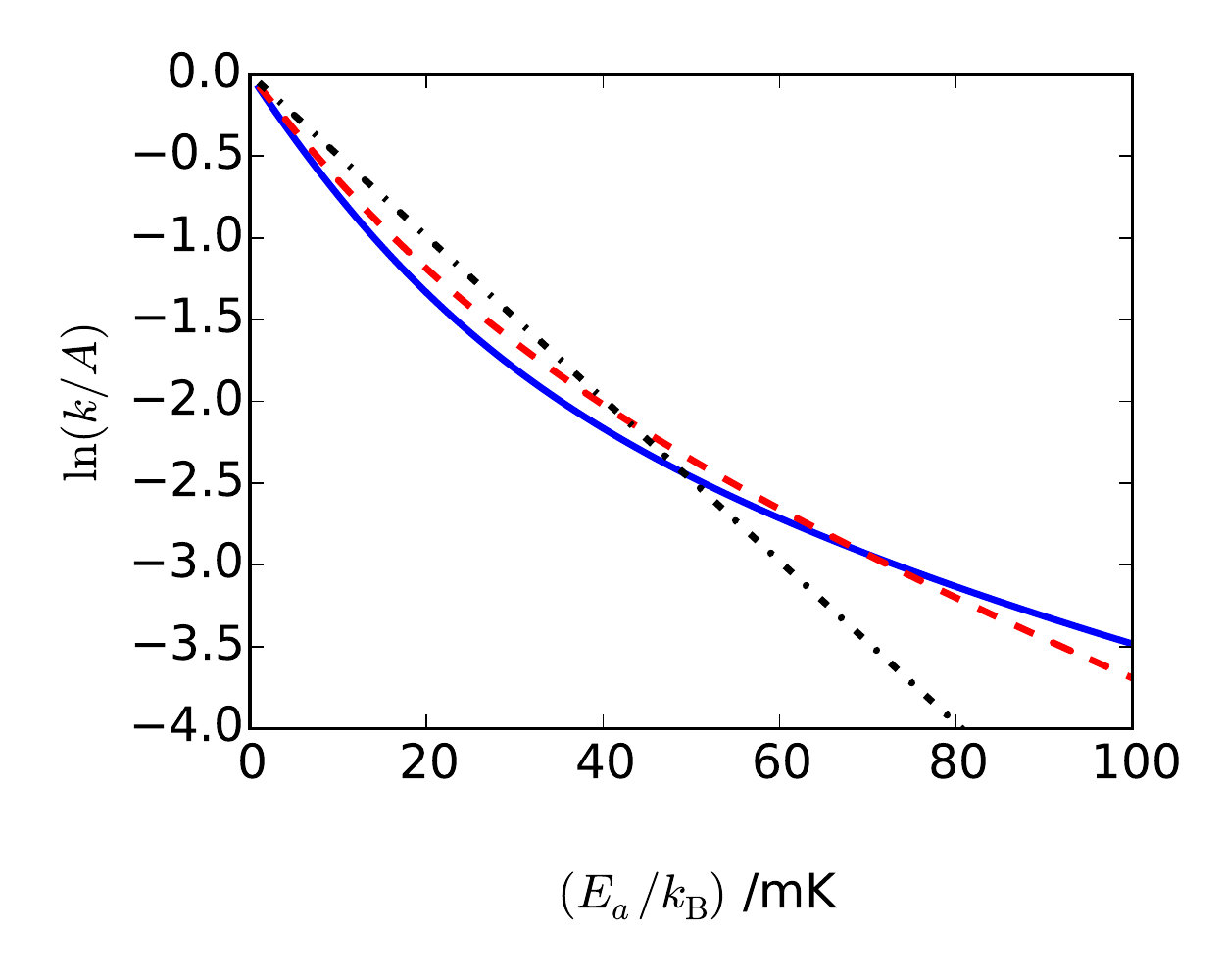}
\caption{Variation of the logarithm of the ratio of the time-averaged reaction rate constant $k$ to the Arrhenius prefactor $A$ as a function of the activation energy $E_a$ averaged over the time-varying temperature for all collision strengths (blue solid line) and for the mean collision strength (red dashed line). The black dash-dotted line indicates the expected behavior for a constant temperature equal to the mean. }
\label{fig:rateConst}
\end{figure}

As a further example of the consequences of a time-dependent temperature, we now turn to the area of determining rate constants for cold chemical reactions \cite{willitsch08a, willitsch12a}. Take the rate constant of a reaction with an activation energy $E_a$ as given by the Arrhenius equation:
\begin{equation}
k = A e^\frac{-E_a}{k_{\text{B}}T(t)},
\end{equation}
where A is a reaction-specific prefactor. Temperature fluctuations lead to a time-averaged rate constant

\begin{equation} \label{eq:timeAveragedRate}
\bar k = \frac{1}{\Delta t} \int ^{t_0 + \Delta t}_{t_0} A e^\frac{-E_a}{k_{\text{B}} T(t)} dt
\end{equation}

differing from the rate constant calculated using a model with $T(t) = \bar T$. Furthermore, additionally averaging over collision velocities and impact parameters leads to further changes in the rate constant.  Fig.~\ref{fig:rateConst} shows the rate constants as a function of the activation energy for all three cases: for a fixed temperature, for a time-averaged temperature, and for a time- and collision-averaged temperature. As expected, for $E_a=0$ (e.g., barrierless Langevin-type ion-neutral processes) the same rate constant is obtained, since in this case $k$ does not depend on the velocity of the ions. However, it can be seen from the figure that for time-varying temperatures, the logarithm of the reaction rate is no longer a linear function of the activation energy due to the periods of time spent at higher and lower temperatures. At higher activation barriers, the rate constant for the temperature-varying model is higher than for the fixed-temperature, suggesting that reactions which would be energetically suppressed at the mean temperature can still occur.

%%%%% MICROMOTION

\subsection{Influence of micromotion}
\label{sec:mm}

The discussion so far focussed on the effect of collisions on the secular motion of the trapped ions. Micromotion is implicitly contained by using a fully time-dependent trap potential in our simulations. However, the effects discussed here are qualitatively different from the micromotion heating of ions in a buffer gas which has been studied in detail in a range of previous publications \cite{major68a,devoe09a,zipkes11a,chen14a,weckesser15a}. In these works, it was shown that ions undergoing collisions with neutral buffer-gas particles exhibit heavy-tailed velocity distributions caused by micromotion disruption and the resulting dissipation of micromotion energy into the secular motion (``RF heating''). Here, we have shown that a non-Maxwellian behavior of the entire ensemble can also emerge because of an energetic collision of a single ion with a neutral particle. This effect does not depend on the specifics of the ion trap and, in particular, does not require the presence of micromotion. It results from a time-dependent ensemble temperature which exists whenever cooling processes are active in combination with infrequent but strong heating effects.

The effects pertaining to the secular energy of the ions discussed here are expected to be less relevant in big crystals in which the energy content of the ions is clearly dominated by micromotion \cite{hall13a}. We expect them, however, to be significant in small crystals or strings centered on the RF null line of the trap, or for larger ensembles in multipole traps for which the micromotion energies are small.

%%%% CONCLUSIONS

\section{Summary and conclusions}

We have implemented an MD model for laser-cooled ions in a RF ion trap which includes a realistic physical representation of laser cooling and the collisions of the ions with background gas molecules. Based on the new MD implementation, we explored constraints on the validity of previously employed friction-force models for laser cooling and show that they lead to a significant overestimation of the cooling rates for energetic ions. We show that infrequent collisions with background gas molecules lead to superstatistical secular-velocity distributions of the ion ensemble independent of the presence of micromotion and that this behavior is tunable through changing laser-cooling parameters. We also show that the effects discussed here can have consequences for the determination of rate constants of cold chemical reactions with small activation barriers.

\begin{acknowledgments}
We acknowledge funding from the Swiss Nanoscience Institute project P1214 and the Swiss National Science Foundation as part of the National Centre of Competence in Research, Quantum Science \& Technology (NCCR-QSIT)
\end{acknowledgments}

\appendix

\section{Averaging collisions over impact parameters}
\label{ap:imp}

To avoid repeating a large number of simulations for all possible impact parameters, we note that for the elastic isotropic collisions of structureless particles considered here, the effects of a collision at a collisional velocity $v_c$ are entirely described by $v_\bot$, the component of $v_c$  normal to the collision surface, while the tangential component plays no role \cite{brey97a}. The overall effect is that a fast glancing collision can impart the same momentum to an ion as a slow head-on collision, leading to the same rise in temperature. $v_\bot$ can be calculated from $v_c,b$ according to:
\begin{equation}
v_\bot = v_c \sqrt{1 - \frac{b^2}{d^2}},
\end{equation}
where $d = \sqrt{\sigma / \pi}$ is the maximum impact parameter. Since a collision along this velocity component is by definition head on (i.e., $b=0$), we have:
\begin{equation}
\Delta T(v_c,b) = \Delta T( v_\bot(v_c,b), 0),
\end{equation}
and so the results obtained in Sec. \ref{section:backgroundGas}, Fig.~\ref{fig:tempVelKick} can be used to describe a collision with arbitrary impact parameter. Eq.~(\ref{eq:deltaTVB}) can then be replaced with an integration over the probability-distribution function $f_{v_\bot}(v_\bot)$,
\begin{equation} \label{eq:deltaTVbot}
\overline{\Delta T} =  \int \Delta T(v_\bot,0) f_{v_\bot}(v_\bot)   \text{d} v_\bot .
\end{equation}
Under the (for the present case very good) assumption that the total velocity of the ion (including micromotion) is much smaller than the velocity of the colliding molecule, $f_{v_\bot}(v_\bot)$ may be derived in closed form as:

\begin{equation} \label{eq:velNormDist}
f(v_\bot) = \frac{2 m v_\bot }{k_{\text{B}} T}  \left( 1 -  \text{erf} \left(v_\bot \sqrt{\frac{m}{2 k_{\text{B}} T}}\right) \right),
\end{equation}
where erf($x$) is the error function. This distribution may then be used in combination with the values of $\Delta T(v_\bot,0)$ taken from simulations to evaluate the integral in Eq.~(\ref{eq:deltaTVbot}).

\section{Time-averaged velocity distribution}\label{ap:timeAvDist}
The form of $T(t)$ given by Eq.(\ref{eq:timeTempSingleExp}) allows for an analytical solution of Eq.~(\ref{eq:velDistIntegrand}) to be found by substituting:
\begin{equation}
\text{d}T  =  -\beta  (T - T_{eq}) \text{d}t,
\end{equation}

so that Eq.~(\ref{eq:velDistIntegrand}) can be written as

\begin{equation} \label{eq:velDistTempIntegrand}
f_v(v) = \frac{-4  \pi v^2}{\beta \Delta t} \int_{T_1}^{T_2}  \frac{\sqrt{\left(\frac{m}{2\pi k_{\text{B}} T}\right)^3} e^{\frac{-mv^2}{2 k_{\text{B}} T}}}{T -T_{eq}} \text{d}T.
\end{equation}

Here, the limits of the integral are given by $T_1 = T_{eq} + \Delta T$ and $T_2 = T_{eq} + \Delta T e^{-\beta \Delta t}$. The same integral may also be found through considering the temperature distribution $f_T(T)$ as follow. In our case, we aim to determine the temperature during the interval of time spanned by $[t_0, t_0 + \Delta t)$. The  distribution of times in this interval is uniform (since, experimentally, we cannot tell at which point a collision has occurred), and equal to $1/ \Delta t \text{d} t$. Eq.~(\ref{eq:timeTempSingleExp}) is used to map this time distribution onto a temperature distribution,

\begin{equation} \label{eq:temperatureDistribution}
f_T(T) = f_t( t(T) )  \frac{\text{d}t}{\text{d}T} =1/(\beta \Delta t (T - T_{eq})),
\end{equation}
where $t(T)$ is the inverse function of Eq.~(\ref{eq:timeTempSingleExp}). The Maxwell-Boltzmann distribution averaged over an arbitrary temperature distribution is given by
\begin{equation}
f_v (v) = \int 4 \pi v^2  \left(  \frac{m}{2 \pi k_\text{B} T} \right)^{\frac{3}{2}} e^{-\frac{m v^2}{2 k_{\text{B}} T}}   f_T(T) \text{d}T
\end{equation}
and substitution of Eq.~(\ref{eq:temperatureDistribution}) into this equation recovers Eq.~(\ref{eq:velDistTempIntegrand}). Using another substitution $u = T_{eq}/T$ gives an integral of the form

\begin{equation}
\int \frac{\sqrt{u} e^{-a u}}{u-1} du,
\end{equation}
where $a = m v^2/(2 k_{\text{B}} T_{eq})$  and the prefactor has been omitted. The integral may be evaluated:

\begin{widetext}
\begin{multline} \label{eq:integrationP1}
\int \frac{\sqrt{u} e^{-a u}}{u-1}  \text{d}u =\frac{e^{-a} \left(-4 \pi  \sqrt{a u} ~\mathcal{T}\left[\sqrt{2 a u},\frac{i}{\sqrt{u}}\right]+i\sqrt{\pi u} \left(e^a-\sqrt{\pi  a}~ \text{erfi}\left(\sqrt{a}\right)\right) \text{erf}\left(\sqrt{a u}\right)+\pi  \sqrt{a u}\right)}{i\sqrt{a u}}
\end{multline}
\end{widetext}

where  $\mathcal{T}(h,a)$  is Owen's T function \cite{owen56a}  and $\mathrm{erfi}(x) = - i~ \mathrm{erf}(i x)$. This result can be used to obtain the velocity distribution plotted in Fig.~\ref{fig:ionVelDist},
\begin{widetext}
\begin{multline}
f_v(v\vert \Delta T) =\frac{4 a e^{-a} \sqrt{a \pi }}{\beta \Delta t \sqrt{v^2}}
 \left[  \left(  \text{erfi}\left(\sqrt{a}\right)  -\frac{e^a}{\sqrt{a \pi }} \right) \text{erf}\left(\sqrt{\frac{m v^2}{2 k_\text{B} T}}\right)-4 i \mathcal{T}\left(\sqrt{\frac{m v^2}{k_\text{B} T}},i  \sqrt{\frac{ T}{T_{eq}}}~\right)\right]^{T=T_2}_{T=T_1},
\end{multline}
\end{widetext}
where $T_1 = T_{eq}+\Delta T(v,b)$ and $T_2 = T_{eq} + \Delta T(v,b) \exp(-\beta \Delta t)$, and the notation $f_v(v\vert \Delta T) $ is used to emphasize that it applies in the time period following a collision resulting in a temperature increase of $\Delta T$.

The distributions of the individual velocity components  $v_i$ with $i=x,y,z$  can be obtained through evaluation of:

\begin{equation} \label{eq:velDistTemp}
f_{v_i}(v_i \vert \Delta T) = -\frac{1}{\beta \Delta t} \int_{T_1}^{T_2}  \frac{\sqrt{\frac{m}{2\pi k_{\text{B}} T}} e^{\frac{-mv_i^2}{2 k_{\text{B}} T}}}{T -T_{eq}} \text{d}T.
\end{equation}
Integration yields:

\begin{widetext}
\begin{equation}
f_{v_i}(v_i \vert \Delta T) = \frac{1}{ \beta \Delta t}\sqrt{\frac{m\pi }{2k_\text{B}  T_{eq}}}   e^{-\frac{m v^2}{2 k_\text{B} T_{eq}}} \left[4 i~\mathcal{T}\left[v \sqrt{\frac{m }{k_\text{B}  T}},-i \sqrt{\frac{T}{T_{eq}}}\right]+\text{erf}\left(v \sqrt{\frac{m }{ 2k_\text{B}  T}}\right) \text{erfi}\left(v \sqrt{\frac{m }{ 2k_\text{B}  T_{eq}}}\right)\right]^{T=T_2}_{T=T_1}
\end{equation}
\end{widetext}

Additionally of interest is the generalized $q_s$ parameter defined as $\frac{\left\langle 1/T^2 \right\rangle}{\left\langle 1/T \right\rangle^2}$ \cite{beck03a}. In the present case, this can be written as:
\begin{equation}
q_s = \beta \Delta t \frac{T_{eq}  (\frac{1}{T_1} - \frac{1}{T_2})    + \log \left(\frac{T_2 (T_1-T_{eq})}{T_1 (T_2-T_{eq})}\right)}{ \log ^2\left(\frac{T_2 (T_1-T_{eq})}{T_1 (T_2-T_{eq})}\right)}.
\end{equation}

\section{Time-averaged Arrhenius rate constant} \label{ap:timeAvRate}
We now derive an expression for the Arrhenius rate constant taking into account temperature fluctuations. Starting from the time-averaged rate constant given by Eq.~(\ref{eq:timeAveragedRate}), we change the integration variable using Eq. (\ref{eq:timeTempSingleExp}):
\begin{equation}
\bar k = \int_{T_1}^{T_2}-\frac{A  e^{-\frac{E_a}{k_{\text{B}} T}}}{\beta \Delta t (T-T_{eq})} \text{d}T,
\end{equation}
Making the substitution $u = T_{eq}/T$ yields:
\begin{equation}
\bar k = \frac{A}{\beta \Delta t}\int_{T=T_1}^{T=T_2} \frac{ e^{-\frac{E_a u}{k_{\text{B}} T_{eq}}}}{ u}-\frac{ e^{-\frac{E_a u}{k_{\text{B}} T_{eq}}}}{ (u-1)} \text{d} u.
\end{equation}
The solution of the integral can be written in terms of the exponential integral function $Ei(x)$ \cite{olver2010a} such that
\begin{multline}
\bar k = \frac{A}{\beta \Delta t} \left[\text{Ei}\left(-\frac{E_a}{k_{\text{B}} T}\right)  \right. \\ \left.  -e^{-\frac{E_a}{k_{\text{B}}  T_{eq}}} \text{Ei}\left(-\frac{E_a}{k_{\text{B}} T }   \left( 1-\frac{T}{T_{eq}}\right)\right)\right]_{T_1}^{T_2}
\end{multline}
which is plotted in Fig.~\ref{fig:rateConst}.

%\bibliography{ian_temp,MainJan15}

%

\end{document}